\documentclass[12pt,twoside]{article}   
\usepackage{epsfig,times}  % 17 dec. 2001: pbs avec times
\usepackage{color}

\begin{document}
\thispagestyle{empty} 

\markboth{{ \sl \hfill Chapter 1 \hfill \ }}
         {{ \sl \hfill Introduction \hfill \ }}

%%%%%%%% Pour changer les valeurs par defaut pour taille figure,
%%%%%%%% sinon au-dela d'une hauteur de 134 mm = 70% on est rejete a la fin
 \renewcommand{\topfraction}{.99}      
 \renewcommand{\bottomfraction}{.99} 
 \renewcommand{\textfraction}{.0}

%%%%% Definitions

\newcommand{\nc}{\newcommand}

\nc{\qI}[1]{\section{{#1}}}
\nc{\qA}[1]{\subsection{{#1}}}
\nc{\qun}[1]{\subsubsection{{#1}}}
\nc{\qa}[1]{\paragraph{{#1}}}

            % Enumerations
\def\qbu{\hfill \par \hskip 6mm $ \bullet $ \hskip 2mm}
\def\qee#1{\hfill \par \hskip 6mm #1 \hskip 2 mm}

\nc{\qfoot}[1]{\footnote{{#1}}}
\def\qL{\hfill \break}
\def\qpar{\vskip 2mm plus 0.2mm minus 0.2mm}
\def\tvi{\vrule height 12pt depth 5pt width 0pt}
\def\qtvi{\vrule height 2pt depth 5pt width 0pt}
\def\qth{\vrule height 15pt depth 0pt width 0pt}
\def\qtb{\vrule height 0pt depth 5pt width 0pt}

\def\qparr{ \vskip 1.0mm plus 0.2mm minus 0.2mm \hangindent=10mm
\hangafter=1}

                % Decale UN paragraphe
                % Attention! La double accolade est vitale, sinon tout le
                % est decale (cf TEX p.199)
                % On peut aller a la ligne avec \qL=\hfill \break
                % Par contre ne supporte pas les lignes blanches
\def\qdec#1{\par {\leftskip=2cm {#1} \par}}

   %% Defs specifiques
\def\qdpt{\partial_t}
\def\qdpx{\partial_x}
\def\qddpt{\partial^{2}_{t^2}}
\def\qddpx{\partial^{2}_{x^2}}
\def\qn#1{\eqno \hbox{(#1)}}
\def\qds{\displaystyle}
\def\qw{\widetilde}
\def\qmax{\mathop{\rm Max}}   % Petit livre Tex (p.167)
\def\qmin{\mathop{\rm Min}}   % Petit livre Tex (p.167)

%%%%% End of definitions

\large

\null
% {\large \it To appear in Physica A}

\color{yellow} 
\hrule height 20mm depth 15mm width 170mm 
\color{black}
\vskip -2.5cm 
\centerline{\bf \Large Impact of sudden mass mortality}
\vskip 4mm
\centerline{\bf \Large on suicides}

\vskip 1.5cm
\centerline{\bf Bertrand M. Roehner $ ^1 $ }
\vskip 4mm
\centerline{\bf Institute for Theoretical and High Energy Physics,
University of Paris 6}

\vskip 0.5cm

{\bf Abstract}\quad We show that a large scale mass mortality
results in increased numbers of suicides.
As a case in point, we consider the influenza epidemic of
October 1918 in the United States. In this month, suicides peaked
at a level of over $ 4\sigma $
(where $ \sigma $ denotes the desaisonalized
standard deviation of the
suicide rate) which means that one would expect
such a jump to occur merely
by chance only once in several centuries. \qL
The mechanism 
that we propose to explain this effect relies on two
steps (i) Mass mortalities break family bonds for instance
between parents and children or husbands and wives.
(ii) Increased numbers of suicides then result from the
well known fact that the severance of family
bonds invariably produces more suicides.

\vskip 1cm
\centerline{4 September 2009}

%\vskip 8mm
%\centerline{\it Preliminary version, comments are welcome}

\vskip 1cm
{\small Key-words: influenza, mass mortality, social bonds, suicide,}
\vskip 1cm

{\small 1: LPTHE, 4 place Jussieu, F-75005 Paris, France. \qL
E-mail: roehner@lpthe.jussieu.fr, Phone: 33 1 44 27 39 16}

\vfill \eject

\qI{Basic observation}

In this paper we show that a sudden mass mortality is
followed within a few weeks by an upsurge of suicides. 
\qpar

What do we mean by the expression ``sudden mass mortality''?
``Sudden'' refers to an event which occurs within one month or
within a few
months. ``Mass mortality'' refers to a monthly mortality 
rate which is at least twice the average rate.\qL
For instance, in September 1918 there were 89,561 deaths in the United 
States which corresponds to a rate of $ 1.10/(\hbox{month}\times
1,000) $ whereas in October 1918 there were 329,400 deaths
corresponding to a rate of  $ 4.12/(\hbox{month}\times 1,000) $;
in other words, as a result of the influenza epidemic,
the death rate was multiplied by 3.7%
\qfoot{All these data are taken from the annual
volumes of the ``Mortality
Statistics'' issued by the Department of Commerce and currently
available online at the website of the ``Centers of Disease Control
and Prevention'' (CDC). These figures in fact only refer
to the part for which 
mortality data were collected (the so-called ``Registration 
Area'') which, in 1918, comprised 79\% of the total population.
We also added the deaths of US soldiers in Europe which, as all
deaths occurring overseas) are not taken into account in the
Mortality volumes.}%
.
\qpar

%%-----------------------------------------------
\begin{table}[htb]

\small

\centerline{\bf Table 1\quad US mortality data for 1918 and 1920
(per 1,000 persons)}

\vskip 5mm
\hrule
\vskip 0.5mm
\hrule
\vskip 2mm

$$ \matrix{
\hbox{}\hfill & \hbox{Influenza} & \hbox{Pneumonia} & 
\hbox{WWI} & \hbox{All} & \hbox{Ratio}\cr
\hbox{}\hfill & \hbox{} & \hbox{and} & 
\hbox{deaths} & \hbox{causes} &  \hbox{to} \cr
\hbox{}\hfill & \hbox{} & \hbox{broncho-} & 
\hbox{} & \hbox{} & \hbox{``normal''} \cr
\qtb
\hbox{}\hfill & \hbox{} & \hbox{pneumonia} & 
\hbox{} & \hbox{} & \hbox{year} \cr
\noalign{\hrule}
\qth
\hbox{\bf ``Normal'' year (1916)}\hfill &  &  & &  & \cr
\hbox{Annual death rate}\hfill & 0.26 & 1.38 &0 & 14.0 & \cr
\hbox{Average monthly death rate}\hfill & 0.02 & 0.11 &0 & 1.17&\cr
\hbox{}\hfill &  &  & &  & \cr
\hbox{\bf 1918}\hfill &  &  & &  & \cr
\hbox{Annual death rate}\hfill & 2.80 & 2.74 &0.95 & 18.7 
& \hbox{\bf 1.3}\cr
\hbox{Monthly death rate in Oct}\hfill
&1.44&0.94&0.57& 4.12& \hbox{\bf 3.5}  \cr
\hbox{}\hfill &  &  & &   &\cr
\hbox{\bf 1920}\hfill &  &  & &  & \cr
\hbox{Annual death rate}\hfill & 0.71 & 1.37 & 0& 13.0 
& \hbox{\bf 0.93}\cr
\qtb
\hbox{Monthly death rate in Feb}\hfill & 0.41 & 0.41 &0 & 
1.91 &\hbox{\bf 1.6}\cr
} $$
\vskip 1.5mm
\small
Notes: Strictly speaking, the data are for the Registration Area only
not for the whole United States, but in 1916 the Registration
Area comprised 71\% of the total population and in 1920 this
percentage was 88\%. During the influenza epidemic
the terminal cause of death was often diagnosed as being
pneumonia or broncho-pneumonia; this is why these two causes
must be taken into account as {\it directly} related to the influenza
outbreak. Regarding the deaths of American soldiers in Europe,
it should be recalled that almost all deaths occurred in 1918 and
that about two-third of the 1918 fatalities occurred in October.
February 1920 constituted a secondary peak in the influenza epidemic
As expected, for mortality events which last only one or two months
the ratio to a ``normal year'' is much higher for monthly than for
annual data. In the latter case the extra-mortality is so-to-say
spread and diluted over the whole year.
\qL
{\it Sources: Mortality Statistics, volumes 1916, 1918, 1920;
Clodfelter p. 785.}
                  % Recapitulation des deces 1918: (S1,48)
\vskip 5mm
\hrule
\vskip 0.5mm
\hrule
\end{table}

%%-----------------------------------------------

What do we mean by an ``upsurge of suicides''? To see this point more
clearly we consider again the example of the United States. Between
September and October 1918 the number of suicides increased 17\% from
752 to 882. However, this increase can hardly be considered as
convincing evidence of a possible effect of the influenza mortality.
Indeed, jumps of this magnitude are fairly common. Just to give
two examples: between February and March 1917 the number of suicides
increased by 17\% and between the same months of 1918 they increased by
12\%. The reason behind the February to March increases is
that suicides (more or less) follow a seasonal pattern with a
minimum in December and a maximum in April or May.
\qpar

So, is the jump which occurred from September to October 1918
completely spurious and irrelevant? Not so. 
\qpar
One obvious question is
whether
the seasonal pattern implies an increase or a decrease 
between September and October.
Over the period 1910-1920 the average seasonal suicide pattern 
reveals a steady fall between May and December; altogether,
the fall is about $ 22\% $. More specifically, 
between September and October the change
is $ -3.6\% $% 
\qfoot{On the contrary between February and May there is a
33\% increase. More specifically, between February and March the 
increase is on average 23\%.}%
.
In this light the 17\% increase 
between September and October 1918 already becomes
more significant. If one takes into account the seasonal
pattern,  the 17\% change once adjusted becomes 
23.6\%. However, this does not
yet allow us to conclude that this change is due to the mass
mortality. It could be a purely random fluctuation.
\qpar

To determine the degree of significance of
this change in a more precise way, we need not only
to compare the Sep-Oct jump in 1918 to the changes over the
same months for other years (which is what we have done by
using the seasonal pattern), we also need to know the
variability of these changes. Depending on the magnitude of
the standard deviation of
these changes, the 1918 jump will appear more or less significant.
This test was carried out for different areas. 
The results are summarized in table 2.

%%-----------------------------------------------
\begin{table}[htb]

\small
\centerline{\bf Table 2\quad How significant is the suicide jump
of October 1918 in the United States}

\vskip 5mm
\hrule
\vskip 0.5mm
\hrule
\vskip 2mm

$$ \matrix{
\hbox{Area} \hfill & \hbox{Jump} &  \hbox{Average}&  \hbox{Standard-}& 
 \hbox{Seasonally} & \hbox{Jump}& \hbox{Ratio}\cr
\hbox{} \hfill & \hbox{in} &  \hbox{Oct. jumps}&  \hbox{deviation}& 
 \hbox{adjusted} & \hbox{expressed}& \hbox{of}\cr
\hbox{} \hfill & \hbox{Oct 1918} &  \hbox{over}&  \hbox{of jumps}& 
 \hbox{jump} & \hbox{in terms}& \hbox{extra-}\cr
\hbox{} \hfill &  &  \hbox{1910-1920}&  \hbox{1910-1920}& 
 \hbox{} & \hbox{of } \sigma& \hbox{suicides}\cr
\hbox{} \hfill & \hbox{} &  \hbox{(except 1918)}&  \hbox{(except 1918)}& 
 \hbox{} & \hbox{}& \hbox{to shock}\cr
\qtb
\hbox{} \hfill & \Delta &  m &  \sigma & 
 \Delta-m & (\Delta -m)/\sigma & \hbox{(per 1,000)}\cr
\noalign{\hrule}
\qth
\hbox{Cities}\hfill      & 70& -7.1& 33.4& 77.1&\hbox{\bf 2.32}& 0.60\cr
\hbox{Rural parts}\hfill & 55&  -5.2& 26.2& 60.2&\hbox{\bf 2.29}& 0.65\cr
\qtb
\hbox{Total area} \hfill & 125 & -12 & 30.1 & 137 & \hbox{\bf 4.6}& 0.62 \cr
\noalign{\hrule}
} $$
\vskip 1.5mm
\small
Notes: October 1918 was the first month of the influenza epidemic.
In most of the states it was during this month that the
losses were largest.
The total area under consideration corresponds to the so-called
``Registration States''. It comprised all states in which death
statistics were duly recorded.
Note that considering monthly changes (rather than monthly 
numbers) makes this analysis independent of possible 
trend modifications (which may be due to
changes in the number of registration states
or other slowly changing factors).\qL
The meaning of ``per 1,000'' in the last
column is that an extra-mortality of 100,000 deaths in a given month
will
result in $ 100\times x $ additional suicides in that month,
where $ x $ denotes the numbers in the last column.
Thus, for cities, 100,000 extra-deaths will result in 60
additional suicides.
\qL
The statistical error bar on $ (\Delta -m)/\sigma $
comes mainly from the confidence interval on
the estimates of $ m $ and is therefore
of the order of $ 1.96/\sqrt{10}=0.6 $ (for a probability 
level of 0.95).\qL
{\it Sources: The primary mortality
data are taken from the annual volumes of
``Mortality Statistics'' published by the US Department of Commerce.}

\vskip 5mm
\hrule
\vskip 0.5mm
\hrule
\end{table}

%%-----------------------------------------------

There are three points of interest in this table.
\qbu Once expressed in terms of the standard deviation of the other
October jumps, the jump $\Delta $
of October 1918 in the whole area is
$ \Delta=4.6 \sigma $; under the assumption of a Gaussian
distribution%
\qfoot{In Roehner (2007, p. 52-53) it was shown that the 
distribution of monthly suicide changes is indeed compatible
with this assumption.}%
, 
this means that in random drawings
a fluctuation of such a
size (or larger) will be observed once in 30,000 drawings 
(see table 3.2 in Roehner (2007). Henceforth,
for the sake of brevity, the ratio $ \Delta/\sigma $ will be
called the $ \sigma $-jump. 
\qbu One may wonder why the $ \sigma $-jump of
the whole area is so much larger than the $ \sigma $-jumps of its
urban and rural parts. The reason is very simple. Whereas the
jump is about twice as high (which is quite natural for an
area which is twice as large) the $ \sigma $ of the total
area is {\it not} twice as high. Why is this so?
By listing the individual changes one is lead to the
observation that most urban and rural changes are 
of comparable magnitude but of opposite sign%
\qfoot{This  is confirmed by the fact that the cross-correlation
of the urban and rural series is -0.51.}%
. 
Thus, by adding the two series one is led to a sum with
a reduced standard deviation. \qL
By generalizing this observation, one is led
to a technique through which it becomes possible to increase the
signal/noise ratio; it will be developed in a subsequent paper.
\qbu As a rule of thumb one should retain that 100,000 extra-deaths
lead to about 60 extra-suicides in the same month.
\qpar

In order to show that these rules hold in a general way, one 
must of course present more than one case. We began with the example 
of the United States in October 1918 because of the conjunction
of two favorable circumstances: (i) A large shock (ii) The existence
of detailed statistics which are readily available through the
Internet. We will shortly present other cases which display the 
same effect but before doing that
we will explain the mechanism which may account for this effect.

\qI{How can one explain this effect?}

There are very few characteristics of suicide which can be 
observed in a fairly universal way that is to say in all
countries and in all times%
\qfoot{For instance one can mention the following facts.
Male suicide rates are higher than
rates for females in western countries whereas in China and
in some parts of India it is the opposite. Around 1890
in the time of Emile Durkheim suicide rates were generally higher 
in cities whereas it the opposite nowadays (e.g. the suicide
rate in Arizona is about twice the rate in New York City).}%
.
One of the most important is the fact that the suicide rates
of married people are lower than the suicide rates of 
bachelors, widows or divorced people. This has been observed
in all countries for which there are reliable data, and over all
times from the 19th century to nowadays. This rule was already 
well-known by Emile Durkheim (1897); a summary of data with
a broad coverage in space and time can be found in Roehner
(2006, p. 257)%
\qfoot{It is important to observe that the rule holds in each
age-group and therefore cannot be attributed to 
age differences, e.g. widowers are on average older than
married people.}%
. 
This rule can be seen as expressing an
equilibrium property. Here, however, we want to consider the
non-equilibrium situation. For the sake of argument let us
consider the specific case of widowers. Before they became 
widowers these people had a suicide rate of, say, 10 per
100,000. Once they have lost their wives their suicide rate
jumps to about 25 per $ 10^5 $. Of course, one would like
to know what is the time-constant $ \tau $ of the transition from
one state to another. Is it a few months or a few years?
To answer this question in a reliable way one would need detailed
data giving the dates of the widowhood and suicide events with an
accuracy of a few weeks. 
\qpar
Anyway, whatever the precise answer to this
question, one can be sure that the sudden
death of
100,000 wives will lead (sooner or later) to 
the suicide of a certain number of widowers which would not have
occurred if those wives had not died. These suicides will
make up a part of the extra-suicides estimated above.
\qpar

So far we have considered the marital bond. This is only one part
of the story, however. Durkheim (1897, p. 59-62) has shown that 
for married people the suicide rate is strongly dependent upon
the size of the family. According to results based on the census of
1886, the suicide rate drops eight-fold
from 40 to 5 per $ 10^5 $
when the number of children increases (less than two-fold) from 1.4 to 2.3%
\qfoot{Such a strong relationship would certainly warrant
additional investigation. Unfortunately, few sources give the number
of children of the persons who commit suicide.}%
.
Now, if we use the same argument as previously, it  can be seen
that the sudden death of 100,000 children will lead (sooner or
later) to the suicide of a certain number of their parents which
would not have occurred if the children had not died. These
suicides will make up another part of the extra-suicides.
\qpar

If this mechanism is accepted at least
as a working hypothesis there is an important implication. 
The fact that a significant number of  extra-suicides occurred
in the {\it same month} as the mass mortality (namely in
October 1918) implies that the time constant $ \tau $ is of
the order of a few weeks rather than a few months or a few years.
Surprising as it may seem, this fact is indeed supported by
evidence collected by researchers who were able to collect
data for the delay between widowhood and suicide events
(see Appendix A).

\qI{Conclusion}

The effect studied in this paper
is fairly weak. It became clearly observable only because 
we carefuly selected the conditions of observation. 
We concentrated our attention on the impact
of the influenza epidemic of 1918 in the United States.
This choice was dictated by two reasons:
(i) This impact was particularly strong 
(ii)  Detailed monthly statistics are available in this case
which put us in a favorable position for carrying out
all required statistical tests.
\qpar

Of course, to validate our present finding
it is essential to show that it holds as well 
for other countries and for
other mass mortalities. This will be done in a subsequent paper. 
\qpar

Broadly speaking, we will use the following strategy.
In the 19th century there were many epidemics which brought about
substantial increases in
mortality. Moreover, the 
recording (and publication) of reliable death statistics 
began around 1850 which means that for several 
countries (e.g. France, the German States, the Scandinavian
countries) it is 
possible to find the {\it monthly} suicide data that are
necessary for this analysis. Naturally, apart from epidemics
there are also other causes of mortality that may
provide good testing opportunities, e.g. the Tokyo-Kanto
earthquake of 1923. In all these cases the first step
(and indeed the main challenge)  is to
get reliable monthly data, possibly at regional level.

\vfill \eject

 \appendix

 \qI{Appendix: The timing of suicides after the disappearance of
a link}

A crucial question 
for the present problem is the delay between the rupture of
a link and the ensuing suicide. To our best knowledge there are
very few papers on this issue. One of the clearest investigations
has been conducted by J. Bojanovsky (1979, 1980) in Germany.
He followed the following procedure.
\qbu For the cities of Heidelberg, Ludwigshaffen and Mannheim 
he was able to get police reports of suicides that occurred between
1971 and 1975. These reports gave the names and marital situation
of the suicides. 
\qbu The names of all persons who were widowed or divorced
at the time of suicide were recorded. To make the subsequent 
work simpler, only persons born in Germany were retained.
\qbu For almost all these persons
the investigators were able to obtain the dates of the
widowhood or suicide from
the official authorities of the ``Land''. 
\qpar

The results are given in Table A1.

%%-----------------------------------------------
\begin{table}[htb]

\small

\centerline{\bf Table A1\quad What is the delay between removal
of the marital bond and suicide?}

\vskip 5mm
\hrule
\vskip 0.5mm
\hrule
\vskip 2mm

$$ \matrix{
\hbox{Delay between dissociation and suicide (month)} \hfill & 0-6 
& 7-12 & 13-24& 25-60 & 61-120 \cr
\qtb
\hbox{Duration of the time interval expressed in semesters} \hfill & 1
& 1 & 2 & 6 & 10 \cr
\noalign{\hrule}
\qth
\hbox{\bf Actual numbers} \hfill &  &  &  &  &  \cr
\hbox{Widowhood, males (43 cases)} \hfill & 17 & 3 & 8 & 8 &  7\cr
\hbox{Divorce, males (55 cases)} \hfill & 17 & 5 & 10 & 19 & 4 \cr
\hbox{Widowhood, females (56 cases)} \hfill & 9 & 5 &10  & 12 & 20 \cr
\hbox{Divorce, females (27 cases)} \hfill & 1 & 1 & 7 & 10 & 8 \cr
\hbox{} \hfill &  &  &  &  & 1 \cr
\hbox{\bf Rates (per month and per 100 suicides)} \hfill &  &  &  &  &  \cr
\hbox{Widowhood, males} \hfill & 40\% & 7\% & 9\% & 3\% & 1.6 \%\cr
\hbox{Divorce, males} \hfill & 31\% & 9\% & 9\% & 6\% & 0.7\% \cr
\hbox{Widowhood, females} \hfill & 16\% & 9\% & 9\% & 3\% & 3\% \cr
\qtb
\hbox{Divorce, females} \hfill & 3.7\% & 3.7\%  & 13\% &6\%  & 3\% \cr
} $$
\vskip 1.5mm
\small
Notes: The figures within parenthesis  in the first column give
the size of the sample. It can be seen that
for widowers the rate in the first semester is about
6 times higher than in the second semester or in the second year. 
It would be interesting to have detailed monthly data for the first
semester but this would require samples at least 10 times larger.
For females, the concentration of the suicides on the first
semester after dissociation is much smaller. This is not surprising
on account of the known fact that females are less affected than males
by a rupture of the marital bond.
The size of the samples given in the first part of the table
shows that one cannot expect too much precision from this investigation.\qL
{\it Sources: Bojanovsky (1979, p. 75, 1980, p. 101).}

\vskip 5mm
\hrule
\vskip 0.5mm
\hrule
\end{table}

%%-----------------------------------------------

As is well-known 
widowed or divorced women are much less prone to suicide 
than men, so we should concentrate our attention on the results
for males. Furthermore, for the purpose of the present paper
we are interested in widowhood rather than in divorce.\qL
It can be seen that for widowers the number of suicides
that occurred in the first semester after the death of the wife
was 5.7 times larger than during the second semester and 4.4 times
larger than during each of the semesters of the second year.
In other words, the effect of a large-scale disappearance of
wives would become almost unobservable after the first semester.
Naturally, for the present investigation one would 
like to know the distribution of suicides
over the successive months of the first semester, but this would
clearly require a much larger sample. On account of the
results in the present paper one would predict that this distribution
has a peak in the first month from which it decreases steadily.
\qpar

From Bojanovsky's results it can be predicted that almost all
the extra suicides in October 1918 which came after widowhood
were committed by men. Let us go a step further and assume that
males are also more affected than females after the loss of a child
(on this point we have no solid indications so far). Under
such assumptions it is possible to derive a prediction which may
be tested.\qL
In 1914, 1915, 1916
the ratios of male to female suicides were: 
3.30, 3.35 and 3.35 respectively (Historical Statistics of the
United States 1975, p. 414). In 
1917, 1918, 1919, for some unknown reason, the ratio fell
to 3.02, 2.93 and 2.71 respectively. The extra-suicides in October
1918 (with respect to September) numbered $ 882-752=130 $. 
Unfortunately, the 1918 volume of ``Mortality Statistics'' does
not give monthly suicide numbers by gender. So 
one must make some assumptions in order to proceed. \qL
If, for the sake of simplicity, we assume that the gender-ratios in
September and October were identical  
for the 752 ``normal'' suicides (we denote it by $ r $)
and that all 130 extra-suicides were committed
by men, then a simple calculation shows that
the gender-ratio in October should be: 
$$ r'= { 752/(1+1/r)+130 \over 752/(1+r) } = r+ { 130 \over 752 }(1+r) $$

For instance, if $ r=2.93 $ one gets: $ r'=3.60 $. Such a change
of 23\% is sufficiently large to be observable. In other words,
it should be possible to check this prediction when detailed
data become available.

\vfill \eject

{\bf References}
\vskip 5mm

\qparr
Bojanovsky (J.) 1979: Wann droht der Selbstmord bei Geschiedenen?
[When after a divorce is the likelihood of committing 
suicide largest?].
Schweizer Archiv f\"ur Neurologie, Neurochirurgie und Psychiatrie
125,1,73-78.

\qparr
Bojanovsky (J.) 1980: Wann droht der Selbstmord bei Verwitweten?
[When after becoming a widower is the likelihood of committing
suicide largest?].
Schweizer Archiv f\"ur Neurologie, Neurochirurgie und Psychiatrie
127,1,99-103.

\qparr
Durkheim (E.) 1897: Le suicide: \'etude de sociologie.
F\'elix Alcan, Paris.\qL
[An English translation appeared in 1951 under the title:
``Suicide, a study in sociology'', Free Press, Glencoe (Illinois)].

\qparr
Clodfelter (M.) 1992: Warfare and armed conflicts. A statistical
reference to casualty and other figures 1618-1991. (2 volumes).
McFarland and Company, Jefferson (North Carolina).

\qparr
Roehner (B.M.) 2007: Driving forces in physical, biological,
and socio-economic phenomena. A network science investigation of
social bonds and interactions.
Cambridge University Press, Cambridge (UK).

\qparr
Roehner (B.M.) 2008: Econophysics: challenges and promises.
An observation-based approach. 
Evolutionary and Institutional Economic Review 4,2,251-266.

\end{document}